# Lower Bounds of Uncertainty of Observations of Macroeconomic Variables and Upper Limits on the Accuracy of Their Forecasts

VICTOR OLKHOV

Independent, Moscow, Russia

victor.olkhov@gmail.com

ORCID: 0000-0003-0944-5113

## Abstract

This paper describes the dependence of the theoretical lower bounds of uncertainty of observations of macroeconomic variables on statistical moments and correlations of random values and volumes of market trades. Any econometric assessments of macroeconomic variables have greater uncertainty. We describe how random values and volumes of trades determine random macroeconomic variables. To predict random macroeconomic variables, one should forecast their probabilities. Upper limits on the accuracy of the forecasts of probabilities of macroeconomic variables, prices, returns, and trades depend on the number of predicted statistical moments. We consider economic obstacles that limit by the first two the number of predicted statistical moments. The accuracy of Gaussian approximations limits the accuracy of any forecasts of probabilities of random macroeconomic variables, prices, returns, and market trades. Any forecasts of macroeconomic variables have uncertainty higher than one determined by predictions of coefficients of variation of random values and volumes of trades.

---

This research received no support, specific grants, or financial assistance from funding agencies in the public, commercial, or nonprofit sectors. We welcome offers of grants, support, and positions.

1. INTRODUCTION

The uncertainties of quantitative estimates of macroeconomic variables and the indeterminacies of their forecasts have been under research for decades. At least since Morgenstern (1950), investigations of "the accuracy of economic observations" fill a long list of references. Cole (1969) studied the effect of the uncertainty of the initial source of economic data collection and the econometric errors on the accuracy of short- or long-term forecasts. Zarnowitz (1967; 1978) provided explicit analysis of the accuracy of short-term macroeconomic forecasts and described the measurements and errors in economists' predictions of changes in aggregate income, output, and the price level. Further, the accuracy of macroeconomic forecasts was studied by Diebold and Mariano (1994), Diebold (2012), Borovička and Hansen (2016), Barrero, Bloom, and Wright (2017), Reif, (2018) and many others. The assessments of the macroeconomic uncertainty using volatility of economic indicators (Jurado, Ludvigson, and Ng, 2015) ties up the uncertainty shocks with business cycles. The cyclical behavior of empirical measures of uncertainty along with business cycles was presented by (Cacciatore and Ravenna, 2020). Bloom (2013) gave a review of the uncertainty's fluctuations problem and discussed the change of uncertainty over time, the possible reasons for variations of uncertainty, the impact of uncertainty fluctuations on short-run investment and hiring, and the consequences of the uncertainty on the recession of 2007-2009. The effects of uncertainty on risk premia and business cycle fluctuations were discussed by Bianchi, Kung, and Tirskikh (2018). The impacts of uncertainty on firms' decisions were studied by Kumar, Gorodnichenko, and Coibion (2022). The studies of the uncertainty of macroeconomic variables cover almost all possible economic and econometric factors and the treatment of macroeconomic uncertainty is a very wide. For example, Gabaix (2011), Acemoglu et al. (2012), and Bloom (2013) consider uncertainty through a description of macroeconomic fluctuations.

We study the uncertainty of economic observations of macroeconomic variables, prices, and returns as the problem stated by Morgenstern (1950): "to form as precise ideas as possible about the accuracy of economic observations." We consider the randomness of values and volumes of market trades as the origin of the lower bounds of the uncertainty of observations of macroeconomic variables and the same time as the cause of upper limits on the accuracy of their forecasts. The randomness of market trades limits any econometric attempts to measure macroeconomic variables "exactly".

The description of random economic processes requires the use of averaging procedures to quantify regular, smooth macroeconomic variables. The duration of the averaging time interval $\varDelta$, to a large extent, determines the uncertainty of econometric valuations of macroeconomic variables and the accuracy of their forecasts. We show how statistical moments and correlations of random values and volumes of trades determine the lower bounds of the uncertainty, the lower bounds of the precision of any econometric valuations of macroeconomic variables during $\varDelta$. The randomness of the values and volumes of market trades is the major economic factor that defines the bounds of uncertainty and the upper limits on the accuracy of forecasts of macroeconomic variables, prices, and returns. We highlight that the same approach describes the bounds of uncertainty and the limits on the accuracy of forecasts of inflation, growth and interest rates, etc. The quantification of these limits requires the development of additional econometric methodologies and valuations.

In Section 2, we discuss a theoretical framework. In Section 3, and App. A we describe the dependence of the lower bounds of uncertainties of observations of macroeconomic variables on the volatilities and correlations of random values and volumes of market trades and propose coefficients of variation as the measure of uncertainty. In Section 4, we discuss economic factors that limit the accuracy of the forecasts of prices, returns, and macroeconomic variables. In App. B., we consider the uncertainty in observations of "complex" macroeconomic variables that depend on different kinds of trades. We assume that all prices are adjusted to the current time.

## 2. Theoretical framework

We consider macroeconomics as a system of agents that perform market deals with various assets, commodities, and services. As agents, we take banks and corporations, plants and factories, households, and shops—all participants in economic and financial transactions at various markets. The sums of similar additive variables of agents such as profits and investment, consumption and supply, etc. define additive macroeconomic variables. The ratios of additive macroeconomic variables define non-additive variables such as prices, inflation, bank rates, GDP rate, etc. The changes of macroeconomic variables completely depend on the changes of agents' variables.

In turn, the changes of agents' variables depend on the values and volumes of market trades. The changes of agents' investment, consumption, and supply during the time interval $\varDelta$ depend on market deals made during $\varDelta$. We consider the economic and financial deals as the only origin of the change of agents' variables. Hence, the values and volumes market trades made during time interval $\Delta$ (without double counting) determine the change of macroeconomic variables.

To quantify the change of macroeconomic variables during $\Delta$ that depend on random values and volumes of market trades one should perform the averaging procedure. The duration of averaging interval $\Delta$ determines the uncertainty of market trades, and finally, the uncertainty of macroeconomic variables. For simplicity, we assume that the interval $\varepsilon$ between two consecutive market deals is a constant that can be equal to or even less than a second. Each market deal at time $t_i$, $t_i-t_{i-1}=\varepsilon$, $i=1,2,...$, changes the corresponding macroeconomic variables. However, the use of market trades with periodicity $\varepsilon \leq 1\ sec$ is almost useless for the description of macroeconomic variables. To describe smooth evolution of macroeconomic variables, one should consider the averaging intervals $\Delta$, which could be equal to weeks, months, or years. Some financial variables, for example, the prices of major stocks, indices, or volatilities, can be considered on a daily basis or even on an hourly basis.

We consider the randomness of market trades during $\Delta$ as the only origin of the uncertainty in observations of macroeconomic variables. We don't study how various factors impact the randomness of the values and volumes of trades. Instead, we describe how the random values and volumes of trades during $\Delta$ determine the lower bounds of uncertainty of observations of macroeconomic variables, prices, and returns and how that limits the accuracy of their forecasts. The consideration of values and volumes of trades, macroeconomic variables, prices, and returns as random variables during the averaging interval $\Delta$ gives a uniform basis for the theoretical description of macroeconomics.

The theoretical assessments of the lower bounds of the uncertainty of macroeconomic variables and the upper limits on the accuracy of macroeconomic forecasts should be supported by econometric estimates. That problem reveals the complementary roles of theoretical economics and econometrics. Actually, the sufficient amount of direct data that is required to calculate the values of macroeconomic variables doesn't exist. There is no sufficient data about all market trades made by each agent during any the interval $\Delta$. Direct econometric assessments of macroeconomic variables as sums of all market trades made by all agents are almost impossible. At that point, our theoretical definitions of macroeconomic variables call for the help of econometric methodologies as a way to estimate the values of macroeconomic variables in the absence of direct data. Econometric methodologies solve these problems and give the approximations of macroeconomic variables that theoretically are determined as sums of the values or volumes of market trades. Econometric methodologies (Fox et al., 2019) give the estimates macroeconomic variables using observable data.

That reveals the duality of the problem of uncertainty of macroeconomic variables. The second part of the problem – the uncertainty of macroeconomic variables as a result of inaccurate econometric data - has been studied deeply (Morgenstern, 1950; Cole,1969; Davidson and MacKinnon 2004; Mills and Patterson 2009; Hansen, 2014; Fox et al. 2019; Ilut and Schneider, 2022). However, the first part of the problem, which reveals the dependence of uncertainty of macroeconomic variables on the randomness of the values and volumes of market trades during the interval $\Delta$, was almost missed. Our article, at least partially, covers this gap and describes the bounds of uncertainty and the limits on the accuracy of forecasts of macroeconomic variables, prices, and returns that depend on the randomness of market trades.

That dependence raises a new, tough challenge for econometric methodologies and calculations. Indeed, current econometrics highly succeeds in estimating macroeconomic variables, which are composed of *sums of $1^{st}$ degrees* of values or volumes of trades during the interval $\Delta$. We call them the $1^{st}$ order variables and denote economic models that describe their evolution as $1^{st}$ order economic theories.

The dependence of macroeconomic variables on random values or volumes of market trades exposes their probalistic nature. We define random macroeconomic variables that depend on random values or volumes of trades. The averages of these random macroeconomic variables coincide with usual values of macroeconomic variables that equal to the sums of the values or volumes of trades during $\Delta$. The random origin of macroeconomic variables explains the treatment of their uncertainties as their volatilities or coefficients of variation. The volatilities of random macroeconomic variables, prices, and returns depend on $2^{nd}$ statistical moments, volatilities and correlations of the values and volumes of trades.

Actually, these variables depend on ***sums of squares*** of values or volumes of trades and we call them $2^{nd}$ order variables. Each average macroeconomic variable of $1^{st}$ order should be complemented by its $2^{nd}$ order macroeconomic volatility. That at least doubles the number of variables that describe macroeconomic evolution. One should take into account the mutual dependence of averages and volatilities of macroeconomic variables that are described by the $1^{st}$ and $2^{nd}$ order variables and that significantly complicates macroeconomic models. We denote the description of mutual dependence of $1^{st}$ and $2^{nd}$ order macroeconomic variables as $2^{nd}$ order economic theories.

To forecasts random macroeconomic variables, prices, returns, and trades one should predict their probabilities. The number of predicted statistical moments determines the accuracy of probability forecast. Predictions of the $2^{nd}$ statistical moments of macroeconomic variables, prices, returns, and trades require use of $2^{nd}$ order economic theory. Currently, $2^{nd}$ order economic theories, the econometric methodologies and econometric assessments of $2^{nd}$ order variables are absent. That means the lack of an economic basis for predictions of $2^{nd}$ order variables and $2^{nd}$ statistical moments. Simply speaking, current forecasts of volatilities of prices, returns, macroeconomic variables, and trades have almost no economic foundation. For many years to come, that limits the accuracy of predictions of their probabilities by Gaussian distributions. One can find more details in Olkhov (2021-2024).

### 3. LOWER BOUNDS OF UNCERTAINTY

To estimate the volatility as a measure of uncertainty of macroeconomic variables, one should consider them as random variables. To define such random variables, at the first step, we consider the changes of macroeconomic variables during $\Delta$ to be equal to the sums of the corresponding variables of economic agents. In turn, the changes of additive variables of agents equal to the sums of the values or volumes of trades made by agents during $\Delta$. Thus, the sums of values or volumes of market trades during $\Delta$ determine the changes of macroeconomic variables during $\Delta$. As example, the changes of macroeconomic investment, credits, and consumption equal to the sums of the investment, credit, and consumption deals (without repeating) made by all agents during $\Delta$. The definitions of some macroeconomic variables use the linear combinations of sums of different market deals. For example, the change of GDP during $\Delta$ equals to the sum of the Value Added (VA) of all agents plus the net export trades (Fox, 2019). To calculate the VA one should sum the linear combinations of trade sales and purchases made by agent during $\Delta$. The particular linear form that defines agents' VA can vary due to different schemes of accounting, business specifics, and tax regulations. Anyway, the linear dependence of VA on market trades made by agents permits us to describe the uncertainty of GDP during $\Delta$ completely in the same way as the uncertainty of such variables as investments, credits, or consumption. Finally, the changes of different additive variables of agents during $\Delta$ can be presented by the linear forms of the sums of various market deals made by agents during $\Delta$. As we show (App. B), the lower bounds of the uncertainties of macroeconomic variables during $\Delta$ are determined by the volatilities, correlations, and coefficients of variation of random values or volumes of market trades.

To illustrate that dependence, as example, we consider the lower bounds of the uncertainty of macroeconomic consumption. Let us assume that during time interval $\Delta$ (3.1), each agent $j$, $j=1,...M$, made purchase deals that resulted in consumption, and there was no doubling. We denote the values $C(t_i;j)$ of the consumption deals made by the agent $j$ at time $t_i$, and the total number of times $t_i$ during $\Delta$ equals to $N$:

$$t - \frac{\Delta}{2} < t_i < t + \frac{\Delta}{2} \quad ; \quad i = 1, ... N \tag{3.1}$$

Macroeconomic consumption $C_m(t)$ during $\Delta$ equals the total value of consumption deals made during $\Delta$. To describe macroeconomic consumption, one should take into account only the total values of consumption deals made by all agents at time $t_i$ during $\Delta$. Let us define the sum $C(t_i)$ of consumption deals $C(t_i;j)$ made by all agents $j=1,..M$ at time $t_i$:

$$C(t_i) = \sum_{j=1}^{M} C(t_i; j) \tag{3.2}$$

The total number of times $t_i$ of consumption deals during $\Delta$ equals to $N$ (3.1). We define the *n-th* statistical moments $C(n)$ of consumption deals during $\Delta$:

$$C(n) = E[C^n(t_i)] = \frac{1}{N} \sum_{i=1}^{N} C^n(t_i) \quad ; \quad n = 1,2,.. \tag{3.3}$$

$$C_\Delta(n) = \sum_{i=1}^{N} C^n(t_i) = N \cdot C(n) = C_m(t) \tag{3.4}$$

Relations (3.3) give the approximations of the *n-th* statistical moments by the finite number $N$ (3.2) of the consumption trades made at time $t_i$ during $\Delta$. The function $C(1)$ (3.3) denotes the average values of the consumption trades during $\Delta$. The function $C_\Delta(1)$ in (3.4) equals to the total value of all consumption deals made by all economic agents during $\Delta$, and it defines the usual macroeconomic consumption $C_m(t)=C_\Delta(1)$ (3.4) during $\Delta$. However, the lack of total data about all consumption deals causes that econometric valuation of macroeconomic consumption $C_m(t)=C_\Delta(1)$ could use completely different econometric methodology to quantify the value of consumption $C_\Delta(1)$ in (3.4) during $\Delta$. We highlight the possible differences between the theoretical definition of a macroeconomic consumption $C_m(t)=C_\Delta(1)$ (3.4), and the definition of macroeconomic consumption given by econometric methodology to measure that variable using the available, observable data.

To estimate the uncertainty of macroeconomic consumption $C_m(t)=C_\Delta(1)$ during $\Delta$, we define macroeconomic consumption as a random variable $x(t_i)$ (3.5):

$$x(t_i) = N \cdot C(t_i) \tag{3.5}$$

In (3.5), we consider the consumption trades $C(t_i)$ that were already made during $\Delta$ (3.1), thus their number $N$ is fixed. We call $x(t_i)$ (3.5) a random macroeconomic consumption during $\Delta$. The mathematical expectation $x(1)$ (3.6) of the random consumption $x(t_i)$ (3.5) takes the form:

$$x(1) = E[x(t_i)] = \frac{1}{N} \sum_{i=1}^{N} x(t_i) = \sum_{i=1}^{N} C(t_i) = C_\Delta(1) = N \cdot C(1) \qquad (3.6)$$

Thus, the average *x(1)* (3.6) of a random consumption *x(t_i)* (3.5) equals usual value of macroeconomic consumption $C_m(t)=C_\Delta(1)$ (3.4) during *Δ*. The definition of macroeconomic consumption as a random variable *x(t_i)* (3.5) permits us to consider its volatility $\sigma_x^2$ (3.7) as the assessment of the uncertainty of macroeconomic consumption $C_m(t)=C_\Delta(1)$ during *Δ*:

$$\sigma_x^2 = E[(x(t_i) - x(1))^2] \qquad (3.7)$$

$$E[x^2(t_i)] = \frac{1}{N} \sum_{i=1}^{N} x^2(t_i) = N \sum_{i=1}^{N} C^2(t_i) = N\, C_\Delta(2) = N^2 C(2) \qquad (3.8)$$

From (3.3 - 3.8), obtain the dependence of volatility $\sigma_x^2$ (3.9) on volatility $\sigma_C^2$ (3.10):

$$\sigma_x^2 = N^2[C(2) - C^2(1)] = N^2 \sigma_C^2 \qquad (3.9)$$

$$\sigma_C^2 = C(2) - C^2(1) \qquad (3.10)$$

In (3.9; 3.10), $\sigma_C^2$ denotes the volatility of the random values *C(t_i)* of consumption trades during *Δ*. The square of the coefficient of variation $\chi_x^2$ (3.11) of a random macroeconomic consumption *x(t_i)* (3.5) equals the square of the coefficient of variation $\chi_C^2$ (3.11) of consumption trade values during *Δ*:

$$\chi_x^2 = \frac{\sigma_x^2}{x^2(1)} = \frac{N^2 \sigma_C^2}{N^2 C^2(1)} = \frac{\sigma_C^2}{C^2(1)} = \chi_C^2 \qquad (3.11)$$

The squares of coefficients of variation $\chi_x^2$ and $\chi_C^2$ (3.11) describe the volatilities of random variables with averages equal to one. We propose them as a measure of the lower bounds of the uncertainty of random macroeconomic variables. Any econometric valuations of macroeconomic variable have the uncertainty higher than $\chi_x^2$ (3.11). The relations (3.11) demonstrate that lower bounds of the uncertainty of the consumption trade values *C(t_i)*, which we measure by $\chi_C^2$ (3.11), coincide with the lower bounds of the uncertainty of macroeconomic consumption $\chi_x^2$ (3.11) during *Δ*. We highlight that the quantifications of the lower bounds of the uncertainty of consumption (3.11) need econometric methodologies to estimate the sums of squares of the values of trades during *Δ*. All of that is absent now.

In App. B., we describe lower bounds of uncertainty of "complex" macroeconomic profits that are determined by different deals.

## 4. UPPER LIMITS ON THE ACCURACY OF FORECASTS

To describe upper limits on the accuracy of macroeconomic forecasts, as an example, we consider consumption. We define a random macroeconomic consumption *x(t_i)* (3.5), which is determined by the random values *C(t_i)* of consumption deals during *Δ* (3.1). To predict a random variable, one should forecast its probability. The accuracy of the forecasts of probability of a random consumption *x(t_i)* (3.5) is determined by the accuracy of the predictions of the probability of random values $C(t_i)$ of consumption deals. The more precise the predictions of probability of consumption deals, the more precise would be the forecasts of macroeconomic consumption. The more statistical moments of a random variable are used for the approximation of the probability of a random variable, the higher the accuracy of the resulting approximation of probability (Shiryaev, 1999; Shreve, 2004). The forecasts of the first two statistical moments define the average and volatility of a random variable and determine Gaussian approximations of probability.

The volatility $\sigma_C^2$ (3.3; 3.10) of the random values $C(t_i)$ of consumption trades depends on the 2$^{nd}$ statistical moment $C(2)$ (3.3), which is determined by the sum of squares of the values of consumption deals during $\varDelta$, and we call it a 2$^{nd}$ order economic variable. As we already discussed, the predictions of $C(2)$ (3.3) require the development of 2$^{nd}$ order economic theory. The quantitative assessments of $C(2)$ (3.3) need econometric methodologies and valuations that are absent now. The assessments of the 3$^{rd}$ or 4$^{th}$ statistical moments require econometric methodologies and quantifications of variables composed by the sums of the 3$^{rd}$ and 4$^{th}$ degrees of the values of consumption deals during $\varDelta$. The predictions of the 3$^{rd}$ or 4$^{th}$ statistical moments require economic theories that model the mutual evolution of variables up to the 3$^{rd}$ or 4$^{th}$ orders. All of that is absent now. That limits the accuracy of predictions of the probabilities of macroeconomic variables, in the best case, by Gaussian distributions.

The predictions of the volatilities of macroeconomic variables are similar to the problem of forecasting the volatilities of prices and returns, which also depend on predictions of the volatilities and correlations of the values and volumes of trades (Olkhov, 2021; 2023b; 2023c; 2024). The dependence of the volatilities of prices, returns, and macroeconomic variables on the sums of squares of values and volumes of different types of market deals ties up the forecasting of their probabilities in a unified puzzle. The upper limits on the accuracy of the predictions of their probabilities are similar and are limited by Gaussian approximations. It should be agreed that there are no economic reasons to imagine that one can predict the probabilities of a particular macroeconomic variable, price, or return with an accuracy that is higher than for others. The upper limits on the accuracy of forecasts depend on the development of 2$^{nd}$ order economic theories. In turn, the development of these theories depends on the creation of econometric methodology to estimate the 2$^{nd}$ order variables. Until then, the upper limits on the accuracy of the forecast of macroeconomic variables, prices, and returns are limited by Gaussian approximations of their probabilities.

In particular, the accuracy of forecasts of macroeconomic consumption is limited by the predictions of its coefficient of variation $\chi_x^2$ (3.11). That problem is equal to predictions of coefficient of variation $\chi_C^2$ of the random values of consumption trades (3.11). The forecasts of coefficient of variation $\chi_C^2$ of the random values of consumption trades determine the upper limits on the accuracy of forecasts of macroeconomic consumption. Any forecasts of macroeconomic consumption have accuracy lower or uncertainty higher than one determined by the predictions of coefficient of variation $\chi_C^2$ of the random values of consumption trades. In turn, the accuracy of predictions of coefficient of variation $\chi_C^2$ of limited by the forecasts of Gaussian approximations of the probability of consumption trades.

## 5. CONCLUSION

The randomness of the values and volumes of market trades determines the uniform basis for description of macroeconomic variables, prices, and returns as random variables, and highlights the probalistic nature of macroeconomic variables and theories. The volatilities and correlations of random values and volumes of trades determine lower bounds of uncertainty of economic observations and upper limits on the accuracy of forecasts of probabilities of macroeconomic variables, prices, and returns. Econometric assessments and description of these limits need the development of $2^{nd}$ order economic theory. One can't prove economically the predictions the probabilities of a particular macroeconomic variable, price, or return with accuracy that is higher than the accuracy of others.

The economic roots of both lower and upper limits depend on $2^{nd}$ order variables determined by the sums of squares of the values or volumes of market trades during $\Delta$. The quantification of $2^{nd}$ order economic variables requires development of econometric methodologies. The current economic theories and econometric methodologies describe the mean values of macroeconomic variables that depend on the sums of the $1^{st}$ degrees of the values or volumes of trades. The mutual description of $1^{st}$ and $2^{nd}$ order macroeconomic variables at least doubles the number of variables and the complexity of their modeling. In simple words, the accuracy of the description of mean values of macroeconomic variables, prices, and returns depends on their $2^{nd}$ statistical moments and volatilities. The more statistical moments that can be predicted, the higher would be the accuracy of the average macroeconomic variables, prices, and returns. The lack of direct data about the values and volumes of market trades made by all economic agents that is required for the valuation of their $2^{nd}$ statistical moments makes direct quantitative assessments impossible. That requires the development of econometric methodologies and theories to quantify and predict the $2^{nd}$ statistical moments, volatilities, and correlations of the random values and volumes of trades. That will permit us to approximate and describe the volatilities of prices, returns, and macroeconomic variables and will support the development of macroeconomic models that describe the mutual evolution of the 1$^{st}$ and 2$^{nd}$ order variables. Only then will forecasts based on Gaussian approximations of probabilities of price, returns, and macroeconomic variables have an economic foundation.

The general origin of the economic complexity for valuation and modeling the 2$^{nd}$ order economic variables establishes common lower limits on the uncertainty of observations of macroeconomic variables, prices, and returns and upper limits on the accuracy of their forecast. However, the descriptions of the general problem could open an opportunity for the development of approximations to "overcome" economic-based limits.

APPENDIX A: COEFFICIENTS OF VARIATION OF PRICES AND RETURNS

We briefly present the results by Olkhov (2022; 2023a; 2023b; 2024) and refer there for details. Let us consider the values $C(t_i)$ and volumes $U(t_i)$ of market trades during the averaging interval $\Delta$ (A.1):

$$t - \frac{\Delta}{2} < t_i < t + \frac{\Delta}{2} \quad ; \quad i = 1, \dots N \quad (A.1)$$

We define price $p(t_i)$ and return $r(t_i,\tau)$ (A.2) for the constant time shift $\tau$ at time $t_i$:

$$C(t_i) = p(t_i)U(t_i) \quad ; \quad r(t_i, \tau) = \frac{p(t_i)}{p(t_i - \tau)} \quad (A.2)$$

One can convert the equation (A.2) into the equation (A.4) on return $r(t_i,\tau)$:

$$C(t_i) = p(t_i)U(t_i) = \frac{p(t_i)}{p(t_i - \tau)} p(t_i - \tau)U(t_i) \quad ; \quad C_o(t_i, \tau) = p(t_i - \tau)U(t_i) \quad (A.3)$$

$$C(t_i) = r(t_i, \tau) C_o(t_i, \tau) \quad (A.4)$$

In (A.3; A.4), $C_o(t_i,\tau)$ denotes the market value of trade volume $U(t_i)$ at past time $t_i$-$\tau$. Equations (A.2) on price and (A.4) on return have the same forms, and that cause the similar forms of their volatilities. The *n-th* statistical moments $C(n)$ of market trade value, volume $U(n)$, and past market value $C_o(n,\tau)$ of random time series with $N$ terms during $\Delta$ (A.1) are estimated similar to (3.3):

$$C(n) = E[C^n(t_i)] \sim \frac{1}{N}\sum_{i=1}^{N} C^n(t_i) \quad (A.5)$$

$$U(n) \sim \frac{1}{N}\sum_{i=1}^{N} U^n(t_i) \quad ; \quad C_o(n, \tau) \sim \frac{1}{N}\sum_{i=1}^{N} C_o{}^n(t_i, \tau) \quad (A.6)$$

Equations (A.2; A.4) mean that one can't define market-based statistical moments of price and return similar to (A.5; A.6). As market-based average price $a(1)$ (A.7), we take the well-known volume weighted average price (VWAP) (Berkowitz et al., 1989; Duffie and Dworczak, 2018). We denote $E_m[..]$ market-based mathematical expectation to differ it from frequency-based mathematical expectation $E[..]$ (3.3; A.5). The 1st price statistical moment or average price $a(1)$ takes the form of VWAP:

$$a(1) = \frac{1}{\sum_{i=1}^{N} U(t_i)} \sum_{i=1}^{N} p(t_i)U(t_i) = \sum_{i=1}^{N} p(t_i)w(t_i; 1) = \frac{C(1)}{U(1)} \quad (A.7)$$

$$w(t_i; 1) = \frac{U(t_i)}{\sum_{i=1}^{N} U(t_i)} \quad ; \quad \sum_{i=1}^{N} w(t_i; 1) = 1 \quad (A.8)$$

Functions $w(t_i;1)$ (A.7; A.8) play the role of weight functions. Markowitz (1952), in his famous work on portfolio choice 37 years earlier than Berkowitz et al. (1989), proposed portfolio return $h(1,\tau)$ (A.9) as weighed by past values. This gives the definition of the average return $h(1,\tau)$ (A.9) in the form similar to VWAP (A.7):

$$h(1, \tau) = \frac{1}{\sum_{i=1}^{N} C_o(t_i,\tau)} \sum_{i=1}^{N} r(t_i, \tau) C_o(t_i, \tau) = \sum_{i=1}^{N} r(t_i, \tau) z(t_i; \tau, 1) = \frac{C(1)}{C_o(1,\tau)} \quad (A.9)$$

$$z(t_i, \tau, 1) = \frac{C_o(t_i,\tau)}{\sum_{i=1}^{N} C_o(t_i,\tau)} \quad ; \quad \sum_{i=1}^{N} z(t_i, \tau, 1) = 1 \tag{A.10}$$

Functions $z(t_i;\tau,1)$ (A.9; A.10) play the role of weight functions similar to $w(t_i;1)$ (A.7; A.8). We define the market-based average of return's time series $r(t_i,\tau)$ (A.2) during $\Delta$ (A.1) as the average return $h(1,\tau)$ (A.9). To justify it, we highlight that one can consider the set of returns $r(t_i,\tau)$ (A.2) during $\Delta$ (A.1) as the returns of the selected portfolio. The identities of the forms of equations (A.2) and (A.4) and the forms of the average price (A.7) and average return (A.9) result in the same forms of their volatilities. For brevity, we present a derivation of market-based volatility of price only. Let us consider the $2^{nd}$ degrees of the trade price equation (A.2):

$$C^2(t_i) = p^2(t_i) U^2(t_i) \tag{A.11}$$

For $m=1,2$ we define the $m$-th statistical moments $p(t;m,2)$ of price similar to (A.7):

$$p(m, 2) = \sum_{i=1}^{N} p^m(t_i) w(t_i; 2) \quad ; \quad w(t_i; 2) = \frac{U^2(t_i)}{\sum_{i=1}^{N} U^2(t_i)} \quad ; \quad \sum_{i=1}^{N} w(t_i; 2) = 1 \tag{A.12}$$

Functions $w(t_i;2)$ (A.12) play the role of weight functions similar to (A.8). To define market-based $2^{nd}$ statistical moment $a(2)$ of price that is consistent with price average $a(1)$ we consider market-based volatility $\sigma_p^2$ of price:

$$a(2) = E_m[p^2(t_i)] \quad ; \quad \sigma_p^2 = E_m\left[\left(p(t_i) - a(1)\right)^2\right] = a(2) - a^2(1) \geq 0 \tag{A.13}$$

We derive volatility $\sigma_p^2$ of price by averaging by weight functions $w(t_i;2)$ (A.12) and from (A.12; A.13) obtain market-based $2^{nd}$ statistical moment of price $a(2)$:

$$\sigma_p^2 = \sum_{i=1}^{N} \left(p(t_i) - a(1)\right)^2 w(t_i; 2) = p(2,2) - 2p(1,2)a(1) + a^2(1) \tag{A.14}$$

Simple transformations of (A.14) give volatility $\sigma_p^2$ and the $2^{nd}$ statistical moment $a(2)$ of price (Olkhov, 2022; 2023a; 2023b; 2024):

$$\sigma_p^2 = \frac{\Omega_C^2 + a^2(1)\Omega_U^2 - 2a(1) corr[CU]}{U(2)} \quad ; \quad a(2) = \frac{C(2) + 2a^2(1)\Omega_U^2 - 2a(1) corr[CU]}{U(2)} \tag{A.15}$$

In (A.15) we use (A.6) and denote volatilities of market trade value $\Omega_C^2$ and volume $\Omega_U^2$:

$$\Omega_C^2 = C(2) - C^2(1) \quad ; \quad \Omega_U^2 = U(2) - U^2(1) \tag{A.16}$$

The correlation $corr[CU]$ (A.17) of trade values and volumes during $\Delta$ takes the form:

$$corr[CU] = E\left[\left(C(t_i) - C(t;1)\right)\left(U(t_i) - U(t;1)\right)\right] = E[C(t_i)U(t_i)] - C(1)U(1) \tag{A.17}$$

The joint average $E[C(t_i)U(t_i)]$ (A.17) of the product of trade value and volume equals:

$$E[C(t_i)U(t_i)] = \frac{1}{N}\sum_{i=1}^{N} C(t_i)U(t_i) \tag{A.18}$$

Market-based volatility $\sigma_r^2(\tau)$ (A.20) and the $2^{nd}$ statistical moment $h(2,\tau)$ (A.21) of return (A.2) have the form similar to (A.15), but past values substitute trade volumes:

$$h(2, \tau) = E_m[r^2(t_i, \tau)] \tag{A.19}$$

$$\sigma_r^2(\tau) = E_m\left[\left(r(t_i,\tau) - h(1,\tau)\right)^2\right] = h(2,\tau) - h^2(1,\tau) \geq 0 \quad (A.20)$$

$$\sigma_r^2(\tau) = \frac{\Omega_C^2 + h^2(1,\tau)\Omega_{Co}^2(\tau) - 2h(1,\tau)corr[CC_o(\tau)]}{C_o(2,\tau)} \quad (A.21)$$

$$h(2,\tau) = \frac{C(2) + 2h^2(1,\tau)\Omega_{Co}^2(\tau) - 2h(1,\tau)corr[CC_o(\tau)]}{C_o(2,\tau)} \quad (A.22)$$

$$\Omega_{Co}^2(\tau) = C_o(2,\tau) - C_o^2(1,\tau) \quad ; \quad E[C(t_i)C_o(t_i,\tau)] = \frac{1}{N}\sum_{i=1}^{N} C(t_i)C_o(t_i,\tau) \quad (A.23)$$

$$corr[CC_o(\tau)] = E[C(t_i)C_o(t_i,\tau)] - C(1)C_o(1,\tau) \quad (A.24)$$

Market-based volatility $\sigma_p^2$ (A.15) of price and volatility $\sigma_r^2(\tau)$ (A.20; A.21) of return define the lower bounds of their uncertainty, which are determined by the volatilities and correlations of market trade values, volumes, and past market values. To compare the lower bounds of uncertainty of different variables, prices, and returns, one should consider the volatilities of random variables with averages equal to one. That is the reason to consider the squares of coefficients of variation of price and return as their lower bounds of uncertainty. Simple transformations of (A.15; A.21) give squares of coefficients of variation of price $\chi_p^2$, return $\chi_r^2(\tau)$, market trade value $\chi_C^2$, volume $\chi_U^2$, past market trade value $\chi_{Co}^2$, and their correlations:

$$\chi_p^2 = \frac{\sigma_p^2}{a^2(1)} \quad ; \quad \chi_p^2 \cdot (1 + \chi_U^2) = \chi_C^2 + \chi_U^2 - 2\frac{corr[CU]}{C(1)U(1)} \quad (A.25)$$

$$\chi_r^2(\tau) = \frac{\sigma_r^2(\tau)}{h^2(1,\tau)} \quad ; \quad \chi_r^2(\tau) \cdot (1 + \chi_{Co}^2(\tau)) = \chi_C^2 + \chi_{Co}^2(\tau) - 2\frac{corr[CC_o(\tau)]}{C(1)C_o(1,\tau)} \quad (A.26)$$

$$\chi_C^2 = \frac{\sigma_C^2}{C^2(1)} \quad ; \quad \chi_U^2 = \frac{\sigma_U^2}{U^2(1)} \quad ; \quad \chi_{Co}^2(\tau) = \frac{\sigma_r^2(\tau)}{C_o^2(1,\tau)} \quad (A.27)$$

The squares of coefficients of variation (A.25-A.27) describe the uncertainty of normalized random variables with an average equal to one. They are very convenient for mutual comparisons of the uncertainty of the market trades, prices, and returns.

A similar approach describes the lower bounds of the uncertainty of other non-additive macroeconomic variables such as bank rates, inflation, GDP growth rates, etc. However, the quantitative assessments of (A.25-A.27) require the development of econometric methodologies, data, and econometric assessments that are absent now.

APPENDIX B: UNCERTAINTY OF "COMPLEX" MACROECONOMIC VARIABLES

Most manuals on probability (Shiryaev, 1999; Shreve, 2004) describe the average and volatility of a random variable composed by the sum of $Q$ random variables. Let us assume that a random variable $a$ takes the form:

$$a = \sum_{q=1}^{Q} \beta(q)a(q) \tag{B.1}$$

Coefficients $\beta(q)$ are not random and random variables $a(q)$, $q=1,..Q$ have their averages $A(q)$, volatilities $\sigma^2(q)$, and correlations $corr(q;k)$:

$$A(q) = E[a(q)] \quad ; \quad \sigma^2(q) = E\left[(a(q) - A(q))^2\right] \tag{B2.}$$

$$corr(q;k) = E[(a(q) - A(q))(a(k) - A(k))] = E[a(q)a(k)] - A(q)A(k) \tag{B.3}$$

The average $A$ and volatility $\sigma_A^2$ of a random variable $a$ (B.1) equal:

$$A = E[a] = \sum_{q=1}^{Q} \beta(q)A(q) \tag{B.4}$$

$$\sigma_A^2 = E[(a - A)^2] = \sum_{q=1}^{Q} \beta^2(q)\sigma^2(q) + 2\sum_{q=1;k>q}^{Q-1} \beta(q)\beta(k)corr(q;k) \tag{B.5}$$

The square of the coefficient of variation $\chi_A^2$ (B.6) of a random variable $a$ (B.1) depends on the squares of the coefficients of variation $\chi^2(q)$ (B.7) of the random variables $a(q)$, their averages $A(q)$ and correlations $corr(q;k)$:

$$\chi_A^2(t) = \frac{\sigma_A^2}{A^2} = \sum_{q=1}^{Q} \Theta(q) \cdot \chi^2(q) + 2\sum_{q=1;k>q}^{Q-1} \Phi(q,k)\, \Psi(t;q) \tag{B.6}$$

$$\chi^2(q) = \frac{\sigma^2(q)}{A^2(q)} \quad ; \quad \Theta(q) = \frac{\beta^2(q)A^2(q)}{A^2} \tag{B.7}$$

$$\Phi(q,k) = \frac{\beta(q)\beta(k)A(q)A(k)}{A^2} \quad ; \quad \Psi(q,k) = \frac{corr(q,k)}{A(q)A(k)} \tag{B.8}$$

$$\sum_{q=1}^{Q} \Theta(q) + 2\sum_{q=1;k>q}^{Q-1} \Phi(q,k) = 1 \tag{B.9}$$

One should take into account that (B.9) doesn't play the role of an averaging because some coefficients $\beta(q)$ and $\Phi(q,k)$ can be negative. Relations (B.4-B.9) present the volatility and coefficient of variation for any "complex" macroeconomic variable that is determined by a linear form of different market trades through the volatilities and coefficients of variations of these trades.

Let us consider macroeconomic profits $Pr$ (B.10) during $\Delta$ in a simple form as a difference between revenues determined as the sums of sales $Sa(t_i)$ and expenses determined as the sums of purchases $Ex(\tau_j)$. Sales $Sa(t_i)$ and purchases $Ex(\tau_j)$ describe deals at different markets, and we denote the time of the deals by different times $t_i$ and $\tau_j$

$$Pr = \sum_{i=1}^{K_S} Sa(t_i) - \sum_{j=1}^{K_E} Ex(\tau_j) = K_S \cdot Sa(1) - K_E \cdot Ex(1) \tag{B.10}$$

In (B.10), $K_S$ and $K_E$ – the total numbers of sales and purchases in economy during $\Delta$. We define a random macroeconomic profits $Pr(t_i;\tau_j)$ (B.11) as:

$$Pr(t_i;\tau_j) = K_S\, Sa(t_i) - K_E\, Ex(\tau_j) \tag{B.11}$$

The average $E[Pr(t_i;\tau_j)]$ (B.12) of a random profits $Pr(t_i;\tau_j)$ (B.11) equals to (B.10):

$$E[Pr(t_i;\tau_j)] = K_S\, E[Sa(t_i)] - K_E\, E[Ex(\tau_j)] = K_S \cdot Sa(1) - K_E \cdot Ex(1) = Pr \tag{B.12}$$

The square of the coefficient of variation $\chi_{Pr}^2$ (B.13) of random profits (B.11):

$$\chi_{Pr}^2 = \chi_{Sa}^2 \cdot \Theta_{Sa} + \chi_{Ex}^2 \cdot \Theta_{Ex} + 2 \cdot \Phi \cdot \Psi \tag{B.13}$$

$$\chi_{Sa}^2 = \frac{\sigma_{Sa}^2}{Sa^2(1)} \;;\; \chi_{Ex}^2 = \frac{\sigma_{Ex}^2}{Ex^2(1)} \;;\; \Theta_{Sa} = \frac{K_S^2\, Sa^2(1)}{Pr^2} \;;\; \Theta_{Ex} = \frac{K_E^2\, Ex^2(1)}{Pr^2} \tag{B.14}$$

In (B.14) $\chi_{Sa}^2$ and $\chi_{Ex}^2$ denote the squares of the coefficients of variation and volatilities $\sigma_{Sa}^2$, $\sigma_{Ex}^2$ of sales $Sa$ and expenses $Ex$ respectively. $Sa(1)$ and $Ex(1)$ (B.15) denote the average values of a single sale and purchase during $\Delta$.

$$Sa(1) = \frac{1}{K_S}\sum_{i=1}^{K_S} Sa(t_i) \;;\; Ex(1) = \frac{1}{K_E}\sum_{j=1}^{K_E} Ex(\tau_j) \tag{B.15}$$

$$\Phi = \frac{K_S K_E\, Sa(1)Ex(1)}{Pr^2} \;;\; \Psi = \frac{corr(S,E)}{Sa(1)\, Ex(1)} \tag{B.16}$$

$$corr(S,E) = E[Sa(t_i)Ex(\tau_j)] - Sa(1)Ex(1) \tag{B.17}$$

If all times $t_i \ne \tau_j$, then the joint mathematical expectation

$$E[Sa(t_i)Ex(\tau_j)] = \frac{1}{K_S K_E}\sum_{i=1}^{K_S} Sa(t_i) \sum_{j=1}^{K_E} Ex(\tau_j) = Sa(1)Ex(1)$$

and correlation $corr(S,E)$ (B.17) between sale and purchase deals equals zero. If $n$ times $t_i$ of sale and times $\tau_i$ purchase deals are the same, for example:

$$t_i = \tau_i \;;\; i = 1,\ldots n \;;\; M_S = K_S - n \ge 0 \;;\; M_E = K_E - n \ge 0 \tag{B18}$$

Then joint mathematical expectation takes the form:

$$E[Sa(t_i)Ex(\tau_j)] = \frac{1}{n}\sum_{i=1}^{n} Sa(t_i)Ex(t_i) + \frac{1}{M_S}\sum_{i=n+1}^{K_S} Sa(t_i)\, \frac{1}{M_E}\sum_{j=n+1}^{K_E} Ex(\tau_j) \tag{B.19}$$

For the case (B.18; B.19), correlation $corr(S,E)$ (B.17) can be not zero.

We propose to use the square of the coefficient of variation $\chi_{Pr}^2$ (B.13) as a lower bound of the uncertainty of any econometric valuations of macroeconomic profits $Pr$. However, the lack of direct data for the calculations of (B.13-B.17) raises a problem of econometric approximations of the squares of the coefficients of variation $\chi_{Pr}^2$, $\chi_{Sa}^2$, $\chi_{Ex}^2$, volatilities $\sigma_{Sa}^2$, $\sigma_{Ex}^2$, and correlations $corr(S,E)$ using available data.